\documentclass{article}

\usepackage{microtype}
\usepackage{graphicx}
\usepackage[font=small, textfont=bf, labelfont=bf, aboveskip=0pt]{caption}
\usepackage{subcaption}
\usepackage{booktabs} %
\usepackage{balance}

\usepackage{hyperref}

\newcommand{\paraf}[1]{\vspace{1mm}\noindent\textbf{#1.}}

\usepackage{xspace}
\newcommand{\NAME}{{\scshape SAFuzz}\xspace}

\usepackage[preprint]{icml2026}

\usepackage{amsmath}
\usepackage{amssymb}
\usepackage{mathtools}
\usepackage{amsthm}

\usepackage[capitalize,noabbrev]{cleveref}

\theoremstyle{plain}

\theoremstyle{definition}

\theoremstyle{remark}

\usepackage[textsize=tiny]{todonotes}

\icmltitlerunning{Learning Where to Fuzz: Semantic-Aware Adaptive Testing for AI-Generated Code}

\begin{document}

\twocolumn[
  \icmltitle{SAFuzz: Semantic-Guided Adaptive Fuzzing for LLM-Generated Code}

  \icmlsetsymbol{equal}{*}

  \begin{icmlauthorlist}
    \icmlauthor{Ziyi Yang}{equal,gt}
    \icmlauthor{Kalit Inani}{equal,gt}
    \icmlauthor{Keshav Kabra}{equal,gt}
    \icmlauthor{Vima Gupta}{gt}
    \icmlauthor{Anand Padmanabha Iyer}{gt}
  \end{icmlauthorlist}

  \icmlaffiliation{gt}{Georgia Institute of Technology, Atlanta, GA, USA}

  \icmlcorrespondingauthor{Anand Iyer}{anand.iyer@gatech.edu}

  \vskip 0.3in
]

\printAffiliationsAndNotice{\icmlEqualContribution}

\begin{abstract}
While AI-coding assistants accelerate software development, current testing frameworks struggle to keep pace with the resulting volume of AI-generated code. Traditional fuzzing techniques often allocate resources uniformly and lack semantic awareness of algorithmic vulnerability patterns, leading to inefficient resource usage and missed vulnerabilities. To address these limitations, we present a hybrid testing framework that leverages LLM-guided adaptive fuzzing to detect algorithmic vulnerabilities efficiently. Our system \NAME integrates prompt-based behavioral diversification, harness generation with problem-specific oracles, and an LLM-based predictor to enable adaptive resource allocation and dynamic early stopping. Evaluating \NAME on CSES algorithmic problems, we improve vulnerability discrimination precision from 77.9\% to 85.7\%, achieve a 1.71$\times$ reduction in time cost compared to SOTA GreenFuzz while maintaining comparable recall. We further observe that combining our approach with existing unit test generation methods yields complementary gains, increasing the bug detection recall from 67.3\% to 79.5\%.
\end{abstract}

\section{Introduction}
Large Language Models (LLMs) have become ubiquitous across diverse domains, from question answering~\cite{yue2025surveylargelanguagemodel, Guo_2025} and retrieval systems~\cite{jiang2025ragosystematicperformanceoptimization} to performing scientific discoveries~\cite{zheng2025automationautonomysurveylarge, Abramson2024}. One of the most popular applications has been in software development, where LLM-powered coding agents such as Claude Code~\cite{anthropic2024claude}, Cursor~\cite{cursor_editor} and GitHub Copilot~\cite{github2025copilot} are changing how developers write code. These tools allow developers to express intent in natural language and thus enable rapid experiments with product features~\cite{productivityllm}. As they become popular, the amount of AI-generated code is increasing in production systems used by millions of users~\cite{quantumrun2026copilot} as well as across several safety/resource critical domains like trading platforms~\cite{polo, hlsrewriter} and flight software~\cite{burke2025robotbuildsrobotsbrain}.

However, current testing approaches have not evolved to match the pace and scale of the code output by LLM-based assistants~\cite{jimenez2024swebench}. Traditional testing approaches such as unit testing, formal verification, and fuzz testing were primarily designed to verify human-written code~\cite{leino2010dafny,manes2021fuzzing}. In addition, AI-generated code can exhibit unexpected behaviors, including algorithmic complexity mismatches, resource exhaustion vulnerabilities, and prompt-induced behavioral variations~\cite{cotroneo2025humanai, liu2024promptinj}. As a result, shipping untested AI-generated code raises risks where seemingly correct programs may fail under adversarial input~\cite{pearce2022copilot}.

Existing testing approaches suffer from several limitations. LLM-based unit test generation~\cite{chen2024chatunitest} can achieve high code coverage and effectively checks functional correctness, but it is inherently limited in detecting runtime failures such as timeouts and overflow behaviors. While formal verification~\cite{leino2010dafny} provides strong correctness guarantees, it typically relies on human experts, and it remains difficult for LLMs to reliably generate effective invariants for complex programs. 

To detect such runtime anomalies without manual overhead, Fuzz Testing~\cite{manes2019art, bohme2017coverage} is a widely adopted. Although effective at identifying crashes and security vulnerabilities, traditional fuzzing is \emph{computationally expensive}. Recent fuzzing frameworks such as ~\cite{greenfuzz2024} reduce the search space by filtering out potential non-vulnerable targets. However, these approaches exhibit limited vulnerability discrimination precision and often misclassify semantically complex yet safe programs. In addition, most fuzzing frameworks allocate uniform time budgets across programs regardless of their risk profiles, \emph{wasting resources} on safe code while ignoring potentially vulnerable targets. Further, automated harness generation typically relies on generic drivers that lack problem-specific constraints~\cite{falsecrashreducer2025} and thus \emph{fails to trigger deeper algorithmic bugs}. 

To address these limitations, we propose \NAME, a semantic-aware adaptive fuzzing framework designed for testing AI-generated algorithmic code. \NAME targets two key challenges in fuzzing AI-generated programs: effective \textbf{fuzz harness generation} and efficient \textbf{fuzz time allocation}. The framework consists of three key stages.
First, \NAME generates diverse prompt variants for each problem to capture prompt-induced behavioral variations in LLM-generated code, exposing more varieties of algorithmic and resource-related vulnerabilities.
Second, \NAME employs an LLM-guided fuzz harness generation agent that extracts problem-specific constraints from the specification and constructs semantic oracles. This enables the detection of failures such as timeouts, overflows, and logical inconsistencies beyond generic crash-based fuzzing.
Third, \NAME integrates LLM-derived semantic features with static code metrics to predict vulnerability risk and adaptively allocate fuzzing resources. 

Unlike GreenFuzz~\cite{greenfuzz2024}, which relies solely static features, \NAME discriminates vulnerabilities with higher precision. Programs with vulnerability scores below a threshold are filtered early, while high-risk targets are prioritized with proportional time budgets, and an early stopping mechanism further reduces unnecessary fuzzing.

\begin{figure}[t]
    \centering
    \includegraphics[width=\linewidth]{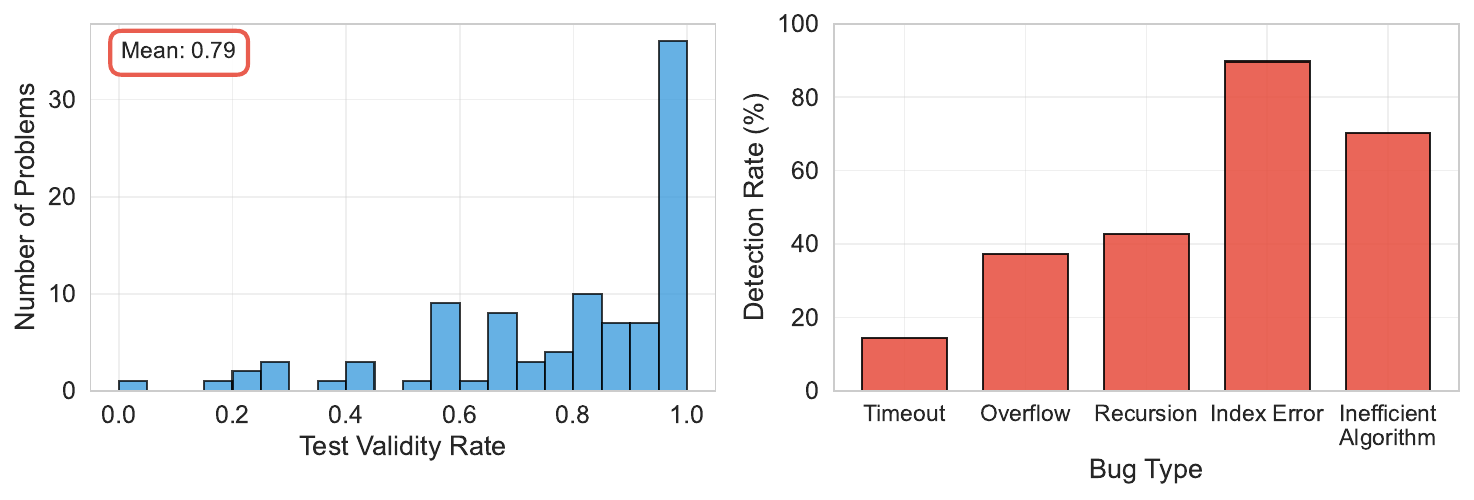}
    \caption{Limitations of ChatUniTest. Left: Validity rate of the generated unit tests across the problems. Right: Bug detection rate across the different bug categories.}
    \label{fig:fuzz_harness_gen_time_breakdown}
\end{figure}

We perform a comprehensive evaluation with a benchmark of 96 CSES algorithmic tasks~\cite{cses2025} as detailed in ~\S \ref{subsec:experimental-setup}, which consists of 1152 program variants with varying difficulty and fuzzing time budgets. 
\NAME improves the vulnerability discrimination precision from 77.9\% to 85.7\% compared to GreenFuzz~\cite{greenfuzz2024}, and improves resource allocation by reducing the total fuzzing time by 1.71$\times$ while maintaining strong recall. By further combining \NAME with unit testing, we enhance the bug detection recall from 67.3\% to 79.5\%. In addition, our fuzz-harness generation agent outputs effective harnesses and maintains linear scalability by employing a bounded retry mechanism that prevents exponential resource costs on complex problems. These results indicate that semantic-aware prioritization can substantially reduce redundant fuzzing effort without sacrificing vulnerability coverage and enables a scalable path toward efficient vulnerability detection.

\section{Background and Motivation}

This section examines the current landscape of testing approaches for AI-generated code and identifies key limitations that motivate our work. We analyze three main categories: LLM-based unit test generation, formal verification approaches, and existing fuzzing techniques.

\subsection{LLM-Based Unit Test Generation}

Unit tests are widely used to evaluate the functional correctness of code in software engineering. Recently, LLM-based unit test generation tools, such as ChatUniTest~\cite{chen2024chatunitest}, have gained popularity. These tools can often generate both test prefixes and test oracles and ensure high coverage of the code under test. Specifically, ChatUniTest introduces a generation-validation-repair pipeline, which includes a feedback loop that iteratively optimizes generated test cases, making sure they are compilable and error-free during runtime. ChatUniTest also implements adaptive focal context generation, which extracts only the most relevant code context to minimize the number of LLM tokens. Overall, ChatUniTest achieves excellent line coverage, but it also often produces invalid tests and fails for complex programs. In \autoref{fig:fuzz_harness_gen_time_breakdown}, we observe an average of 21\% invalid tests generated per problem (left figure). By invalid, we mean that the expected output of a function (asserted in its unit test) does not match the actual output for the given input. Moreover, these tests fail to capture timeouts, memory overflows, or algorithmic complexity mismatches (right figure), making it difficult to detect resource-bounded vulnerabilities.

\subsection{Formal Verification and SMT-based Analysis}

Formal verification mathematically proves the program's correctness by checking if it adheres to the provided formal specification. Tools like Dafny~\cite{leino2010dafny} leverage SMT solvers to automatically verify programs annotated with preconditions, postconditions, and loop invariants. Overall, it provides the strongest correctness guarantees in principle.

However, formally verifying AI-generated code at scale is challenging. It requires precise specifications from domain experts. Recent work explored using LLMs to generate loop invariants~\cite{akhond2025llmloopinvariant, wei2025invbench}, but LLMs struggle with good invariant generation and repairing incorrect invariants. Verification also requires manual effort, iterative refinement, and is computationally expensive, requiring minutes per program. This makes it infeasible to scale for frequently verifying AI-generated code.

\subsection{LLM-Guided Fuzzing}

Fuzzing~\cite{manes2019art} is a popular testing technique that checks programs with randomly generated inputs to discover security vulnerabilities. Traditional fuzzing approaches target programs from their public entry points, but struggle with low coverage on deeply nested code paths. Coverage-guided greybox fuzzers like AFL~\cite{afl2020} address this by using code coverage feedback to guide input mutation, but still face challenges reaching functions deep in the call graph.

Fuzzing relies on good harnesses for targeted testing. There have been recent improvements using LLMs to automate fuzz harness generation. OSS-Fuzz-Gen~\cite{ossfuzzgen2024} uses LLMs in a multi-agent system to automatically generate fuzz harnesses for target functions. Their system comprises harness generation, refinement, coverage analysis, and feedback loops to iteratively improve harnesses. While this approach reduces manual effort, the generated harnesses can report false positive crashes \cite{falsecrashreducer2025} and fail to account for the problem-specific constraints. This leads to an exploding input fuzzing search space and detection of general crashes. 

Another problem with traditional fuzzing is resource allocation and determining the stopping criterion. GreenFuzz~\cite{greenfuzz2024} tries to address this by using machine learning to predict vulnerable functions and stop fuzzing when the coverage of the predicted vulnerable code saturates. They extract features from static analysis tools and software metrics and use the trained classifier to predict vulnerability probability and filter programs. In their evaluation, they terminate campaigns 6-12 hours earlier and miss fewer than 0.5 bugs on average. Their vulnerability predictor achieved ROC-AUC scores of 0.8 (0.827 in evaluation on our dataset). However, GreenFuzz's approach has several limitations. It relies solely on static features and cannot capture algorithmic properties. As shown in \autoref{fig:discrimination}, we can see that there is still a gap in the quality of vulnerability discrimination of Green Fuzz. It is not able to filter out most of the non-vulnerable targets at lower thresholds. As a result, resources are still spent on testing safe code. Moreover, it uses a fixed saturation window and does not adapt fuzzing time based on vulnerability probability, allocating equal time to all targets that pass the vulnerability threshold.

\section{System Design}

To address the limitations identified in the previous work and efficiently detect vulnerabilities in AI-generated code, \NAME combines LLM-guided semantic analysis with traditional fuzzing to achieve intelligent resource allocation. The system takes the natural language descriptions of a set of programming problems as input and assesses the coding agent-generated solutions for vulnerabilities. The pipeline consists of three main stages: (1) Prompt Variant Generation creates diverse formulations of each problem to simulate real-world usage patterns, (2) LLM-Based Fuzz Harness Generation produces problem-specific test harnesses with semantic oracles, (3) Vulnerability Prediction and Adaptive Allocation uses a hybrid ML model to predict vulnerability risk and intelligently distribute fuzzing resources.

\begin{figure*}[t]
    \centering
    \includegraphics[
        width=0.95\textwidth,
        height=0.3\textheight,
        keepaspectratio
    ]{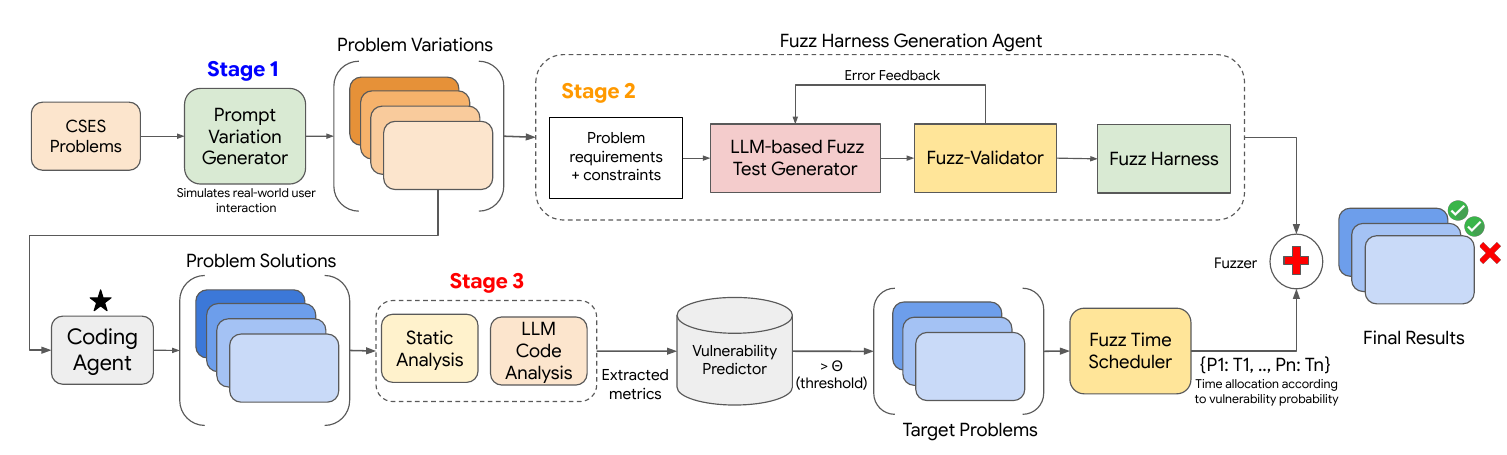}
    \caption{Overview of \NAME. This pipeline has three stages. Stage 1 generates diverse prompt variations to simulate real-world user interactions. Stage 2 involves generation of problem-specific fuzz harnesses with semantic oracles. In Stage 3, a hybrid vulnerability predictor analyzes static and LLM-guided code features to estimate risk, filters out non-vulnerable code and intelligently allocates fuzzing time budgets. Finally, the remaining programs are fuzzed against their harnesses according to their allocated budget. Overall, \NAME prioritizes fuzzing high-risk targets to efficiently detect vulnerabilities.}
    \label{fig:discrimination}
\end{figure*}

\subsection{Prompt Variant Generation}

Previous research has shown that semantically equivalent prompts can trigger different code generation behaviors in LLMs~\cite{sarker2024syntacticrobustnessllmbasedcode, paleyes2025promptvariabilityeffectsllm, chen2025nlperturbator}, which leads to solutions with different types of vulnerabilities. To explore this finding and simulate diverse real-world user interactions, we generated multiple prompt variants for each problem.

Our variant generation strategy is detailed in \autoref{tab:prompt_variants}. For each variant, we preserve the core problem semantics while prompting with various styles, which instruct coding agents to produce different solutions. Each solution is evaluated against the official CSES test suite to mark whether it is buggy. The buggy rate in the table indicates that prompting the same problem in different ways results in variation in vulnerabilities.

\begin{table}[t!]
  \centering
  \small
  \caption{Prompt variants generated for each problem. Each problem generates 12 variants: 1 original, 5 semantic variations, and 6 buggy variations with intentionally injected vulnerability instructions.}
  \label{tab:prompt_variants}
  \begin{tabular}{@{}lp{0.37\linewidth}r@{}}
    \toprule
    \textbf{Variant Name} & \textbf{Description} & \textbf{\% Buggy} \\
    \midrule
    \multicolumn{3}{@{}l@{}}{\textit{Original \& Semantic Variations}} \\
    Original & Original problem format & 29.17 \\
    Overflow Emphasis & Highlights edge cases with large numbers & 27.08 \\
    Reordered Presentation & Reorders constraints and examples first & 29.17 \\
    Examples Only & Minimal format relying on examples & 35.42 \\
    Iterative Approach & Explicitly requests iterative loops & 29.17 \\
    Edge Case Focus & Emphasizes boundary value testing & 28.12 \\
    \midrule
    \multicolumn{3}{@{}l@{}}{\textit{Buggy Variations (Intentional Vulnerabilities)}} \\
    Integer Overflow & Forces int instead of long data types & 62.50 \\
    Timeout/Inefficient & Suggests inefficient algorithms & 93.75 \\
    Heavy Recursion & Encourages deep recursion patterns & 94.79 \\
    Array Indexing & Induces off-by-one errors & 91.67 \\
    Greedy Implementation & Suggests wrong algorithmic approach & 92.71 \\
    Incorrect Logic & Introduces subtle logical errors & 86.46 \\
    \bottomrule
  \end{tabular}
\end{table}

\subsection{LLM-Based Fuzz Harness Generation}

\paraf{Harness Generator Agent Design}
Our fuzz harness generator employs structured prompt engineering to guide an LLM agent in generating Jazzer~\cite{jazzer2021} fuzz tests. The generator receives three inputs: the problem description, the solution code, and the extracted \texttt{solve()} target function signature. The LLM performs semantic analysis to understand algorithmic properties and potential failure scenarios, focusing on two primary tasks. First, \textbf{Constraint Extraction} involves parsing problem descriptions to extract numerical bounds (e.g., $1 \leq n \leq 10^6$), type requirements, and structural constraints. These constraints ensure the generator respects problem-specific limits, such as node and edge counts in graph problems, to avoid exploding the search space. Second, \textbf{Weighted Input Generation} replaces uniform random sampling with weighted distributions to bias toward stress-inducing inputs. For example, it targets maximum recursion depth for recursion-heavy problems (DFS, backtracking), or maximum input sizes for memory-intensive tasks (involving dynamic programming).

\paraf{Oracle Generation}
Generic fuzz harnesses often generate inputs without considering problem constraints, wasting resources on invalid inputs. We address this by generating problem-specific oracles. We implement four oracle types: (1) a timeout oracle that runs solutions in a separate monitoring thread to detect infinite loops, (2) 
a crash oracle for memory violations like null pointer dereferences and array index violations, (3) a determinism oracle to ensure consistent outputs for identical inputs, (4) an overflow oracle that checks for incorrect type choices (for example, operations on large positive numbers that yield a negative value can indicate overflow issues.)

\paraf{Validation Loop}
LLM-generated code can contain syntax errors, type mismatches, or incorrect API usage. Drawing inspiration from OSS-Fuzz-Gen~\cite{ossfuzzgen2024}, we implement a two-stage validation loop with compilation-driven feedback to improve harness quality.

When fuzzing on the generated harness, the agent attempts to compile with the dependencies of the fuzzing framework. The compilation checker captures error messages and parses them to identify common issues like missing dependencies, incompatibility between harness and target function, inappropriate usage of fuzzer API, etc. 
In case of failures, it uses a retry mechanism that includes the original problem, the previously generated harness code, and the compilation error output. This feedback guides the LLM perform targeted fixes rather than regenerating from scratch.

\subsection{Vulnerability Prediction and Adaptive Allocation}
Traditional fuzzing allocates uniform time budgets to all targets, wasting resources on safe code while missing complex vulnerabilities in high-risk programs. \NAME addresses this through ML-based risk prediction and proportional resource allocation, utilizing a three-part pipeline: feature extraction, vulnerability prediction, and adaptive time allocation with early stopping.

\paraf{Feature Extraction}
To capture both structural complexity and algorithmic behavior, we extract 14 features categorized into two distinct groups. First, we compute \textbf{Static Analysis Features} by building upon GreenFuzz~\cite{greenfuzz2024} to calculate six complexity metrics using SciTools Understand~\cite{understand2025} and custom analyzers. These include total lines of code (LOC), cyclomatic complexity, cognitive complexity, and three Halstead metrics (volume, difficulty, effort). These metrics provide a baseline of code size and structural complexity. For example, high cyclomatic complexity often indicates deeply nested control flow that may contain edge case bugs. Second, we introduce \textbf{LLM Semantic Features} because static metrics cannot distinguish between algorithmically safe and unsafe code. A program with low cyclomatic complexity may still implement O($n^2$) logic where constraints require O($n \log n$), or use \texttt{int} datatype where \texttt{long} is needed to avoid overflow. To address this, we prompt an LLM to analyze each program against its problem specification as a human would do, and assign risk scores (0-10) across eight vulnerability categories shown in Table~\autoref{tab:llm_features}. Our LLM prompt includes the problem statement and agent-generated code, along with a detailed scoring rubric for each category. For example, the integer overflow rubric specifies: 0 points for correct use of \texttt{long}, 3 points for potential overflow in edge cases, 7 points for \texttt{int} used in intermediate calculations that can overflow given constraints, and 10 points for definite overflow like \texttt{int*int} multiplication without casting. This prompting approach with a guided rubric reduces LLM hallucination and ensures consistent scoring across problems. These semantic features complement static metrics by reasoning about problem-specific risks; for instance, two programs with identical cyclomatic complexity may receive vastly different timeout risk scores if one uses exponential recursion while the other uses iteration.

\begin{table}[t]
  \centering
  \small
  \caption{LLM semantic vulnerability features (scored 0-10)}
  \label{tab:llm_features}
  \begin{tabular}{@{}lp{0.58\linewidth}@{}}
    \toprule
    \textbf{Category} & \textbf{Detection Focus} \\
    \midrule
    Array bounds risk & Unchecked array accesses, missing bounds validation \\
    Integer overflow & \texttt{int} where \texttt{long} needed (e.g., summing $n=10^6$ values) \\
    Null pointer risk & Dereferencing without null checks, uninitialized variables \\
    Edge case handling & Empty input, max constraint values \\
    Off-by-one error & For a 0-indexed array of size n, accessing the \texttt{n+1}th element \\
    Input validation & Validation against problem constraints \\
    Logic error risk & Algorithm correctness (e.g., greedy where DP is needed) \\
    Timeout risk & Algorithmic complexity vs constraints (e.g., O($n^2$) with $n=10^6$) \\
    \bottomrule
  \end{tabular}
\end{table}

\paraf{Vulnerability Predictor Model}
We train a Random Forest~\cite{breiman2001random} classifier to predict the probability of vulnerabilities from the 14 extracted features. Among different classification approaches such as logistic regression, SVM, and neural network, our experiments show that Random Forest serves the best performance and simplicity.

We used 100 decision trees with a maximum depth of 10 and balanced class weights to handle the imbalanced distribution of buggy versus clean code in our dataset. The model is trained on a 50-50 stratified split of our CSES dataset with 10-fold stratified cross-validation to ensure robust performance estimates.
For each program $i$, the model outputs vulnerability probability $p_i \in [0,1]$ representing the estimated likelihood of containing vulnerable bugs.

\paraf{Adaptive Time Allocation}
Given the vulnerability probabilities for all programs, we allocate fuzzing time to maximize the discovery of bugs under resource constraints by utilizing two primary mechanisms. First, we apply \textbf{Threshold-Based Filtering}, where programs with a vulnerability probability below a threshold $\theta$ are excluded from fuzzing. Let $S = \{i : p_i \geq \theta\}$ denote the set of included programs. This filtering strategy aggressively reduces the fuzzing search space by focusing resources on high-risk targets; for instance, by using $\theta = 0.3$, we achieve 85.7\% precision while retaining 90.2\% of true bugs. Second, we implement \textbf{Proportional Time Allocation} for the included programs, allocating time proportional to their vulnerability probability:
\begin{equation}
t_i = \begin{cases}
\frac{p_i}{\sum_{j \in S} p_j} \cdot T_{\text{budget}} & \text{if } i \in S \\
0 & \text{otherwise}
\end{cases}
\end{equation}
where $T_{\text{budget}}$ is the total budget provided by the user. This ensures that the total fuzzing time scales linearly with the number of included programs while allocating proportionally more time to higher-risk targets within that set.

\textit{Example Calculation:} Consider three programs (P1, P2 and P3) with probabilities $p_1=0.8$, $p_2=0.4$, $p_3=0.2$ and $\theta=0.3$,  $T_{\text{budget}} = 120$s. P3 is filtered ($p_3 < \theta$), leaving $S=\{1,2\}$. As per our allocation scheme, P1 receives $t_1 = 80$s, and P2 receives $t_2 = 40$s for fuzzing.

\paraf{Early-Stopping Scheduler}
Studies on fuzzing have shown that coverage grows rapidly initially, and then plateaus as the fuzzer exhausts reachable states~\cite{bohme2020fuzzing}. Continuing to fuzz after saturation wastes time, which should have been better spent on testing other targets. To address this, our scheduler implements dynamic early stopping based on coverage stagnation detection.

During fuzzing, we parse Jazzer's real-time output to extract combined coverage metrics (edge coverage + feature coverage) at regular intervals. We maintain a timestamp of the last coverage increase for each target. When current coverage remains unchanged for a saturation window $w$, the scheduler preempts the running fuzz instance and schedules the next one.

\section{Implementation}

We implement \NAME in Python using scikit-learn~\cite{pedregosa2011scikit} for ML model training, Jazzer~\cite{jazzer2021} for coverage-guided fuzzing, and DeepInfra~\cite{deepinfra2025} for LLM inference. Our code is organized into modules, including LLM clients, fuzz harness generator, static analysis extractors, vulnerability predictor, and adaptive fuzzing orchestrator.

\paraf{LLM Infrastructure} We use DeepInfra API endpoints for LLM inference. For all experiments, we choose Qwen/Qwen3-Coder-480B-A35B-Instruct-Turbo~\cite{qwen3technicalreport} for its strong code understanding capabilities.

\paraf{Fuzzing Framework} We use Jazzer~\cite{jazzer2021}, a coverage-guided fuzzer built on libFuzzer~\cite{libfuzzer2016} with JVM integration. Our adaptive scheduler parses Jazzer’s output dynamically using regex to extract coverage metrics and detect saturation. Crashes are saved with SHA-1 hashed filenames for automatic de-duplication.

\paraf{Feature Extraction Pipeline} Static metrics are extracted using SciTools Understand~\cite{understand2025} for LOC, cyclomatic complexity, and Halstead metrics, and a custom javalang-based analyzer for cognitive complexity. LLM semantic features are extracted via inference through DeepInfra with structured prompts.

\paraf{Hardware and Deployment} We run our experiments on
a Linux machine with an Intel Core i7-10750H (6-core) and 16GB RAM. We run each Jazzer instance sequentially.

\section{Evaluation}

We evaluate \NAME on CSES algorithmic problems to answer five key questions: (1) How does our hybrid model compare to static-only approaches in vulnerability discrimination? (2) What is the end-to-end bug detection performance and resource savings versus baselines? (3) How does the time budget allocation affect the number of bugs detected? (4) How much can we improve bug detection recall by combining \NAME with unit test generation? (5) How does the fuzz harness generation scale?

\subsection{Experimental Setup}
\label{subsec:experimental-setup}
\paraf{Dataset}
Our dataset consists of 96 algorithmic problems from the CSES benchmark~\cite{cses2025}. The problems cover a wide range of topics, including sorting, greedy algorithms, dynamic programming, graphs, trees, range queries, and mathematics. We manually write an optimal solution (to serve as ground truth) for each problem and verify it against the official test suite.

We generate 12 variants of each problem using our prompt variant generator. An LLM coding agent then generates solutions for all variants, resulting in a total of 1,152  solutions. We run these solutions against official test suites and label each solution as either buggy or clean. For the vulnerability predictor, we use a 50-50 stratified train-test split. The test set contains 336 buggy and 240 clean solutions. Bug categories include timeouts due to algorithmic complexity violations, integer overflows, off-by-one errors, stack overflows, memory out-of-bounds accesses, logic errors etc.

\paraf{Baselines}
We compare against four approaches: (1) ChatUniTest~\cite{chen2024chatunitest}, a state-of-the-art LLM-based unit test generator, (2) Fixed-time baseline fuzzing that allocates equal fuzzing time to all fuzz targets, (3) GreenFuzz~\cite{greenfuzz2024}, an ML-based approach using only static features for vulnerability prediction, and (4) \NAME. All baselines run on identical hardware to ensure fair comparison. Moreover, all fuzzing based approaches use the same harnesses generated by our generation agent (Stage 2).

\paraf{Evaluation Metrics}
We measure the number of bugs caught, recall (\% of actual bugs detected), accuracy (\% of solutions correctly identified as buggy or clean), and time cost (total minutes for complete fuzzing campaign).

\begin{figure}[t]
    \centering
    \includegraphics[width=0.8\columnwidth]{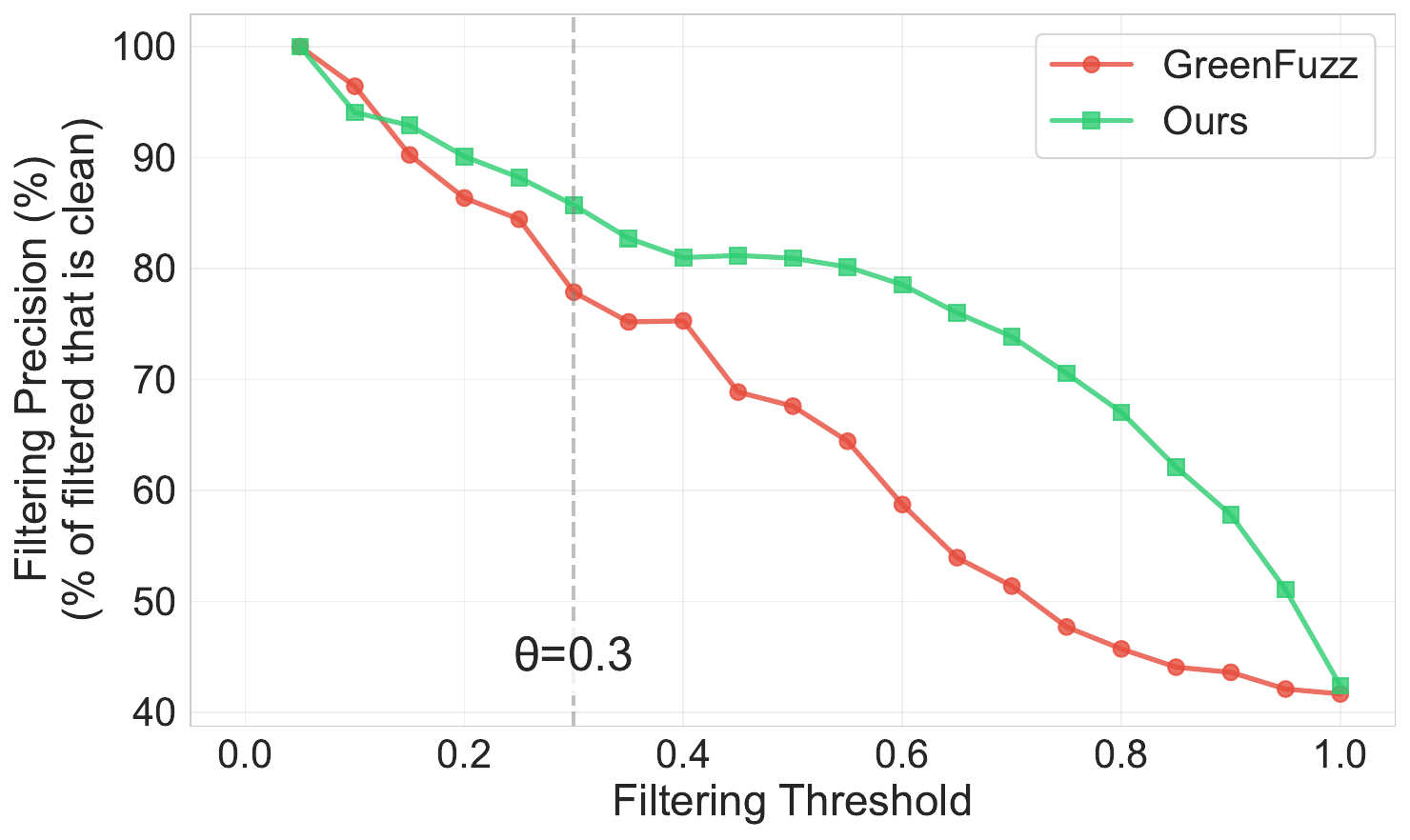}
    \caption{Precision comparison across filtering thresholds. \NAME maintains higher  precision across the threshold range}
    \label{fig:threshold_filtering_precision}
\end{figure}

\subsection{Model Discrimination Quality (RQ1)}

\NAME significantly improves vulnerability discrimination over GreenFuzz's model using only static features. We achieve a mean cross-validation ROC-AUC of 0.943 and outperform the baseline (0.827 ROC-AUC). \autoref{fig:threshold_filtering_precision} shows filtering precision across different vulnerability thresholds. At $\theta=0.3$, GreenFuzz filters 113 programs (88 clean, 25 buggy), achieving 77.9\% precision, while \NAME filters 231 programs (198 clean, 33 buggy) with 85.7\% precision.

\autoref{fig:discrimination} further visualizes this discrimination capability. Points far below the diagonal "Equal Filter Rate" line indicate better performance, that is, filtering more clean code and retaining buggy code for fuzzing. At $\theta=0.3$, \NAME filters 82.5\% of clean code while filtering only 9.8\% of buggy code, compared to GreenFuzz, which filters 36.7\% clean and 7.4\% buggy. By filtering more than twice as many programs (231 vs 113) while missing only 8 additional bugs, our model allows for aggressive resource savings for fuzzing without significantly affecting bug detection.

\begin{figure}[t]
    \centering
    \includegraphics[width=0.8\columnwidth]{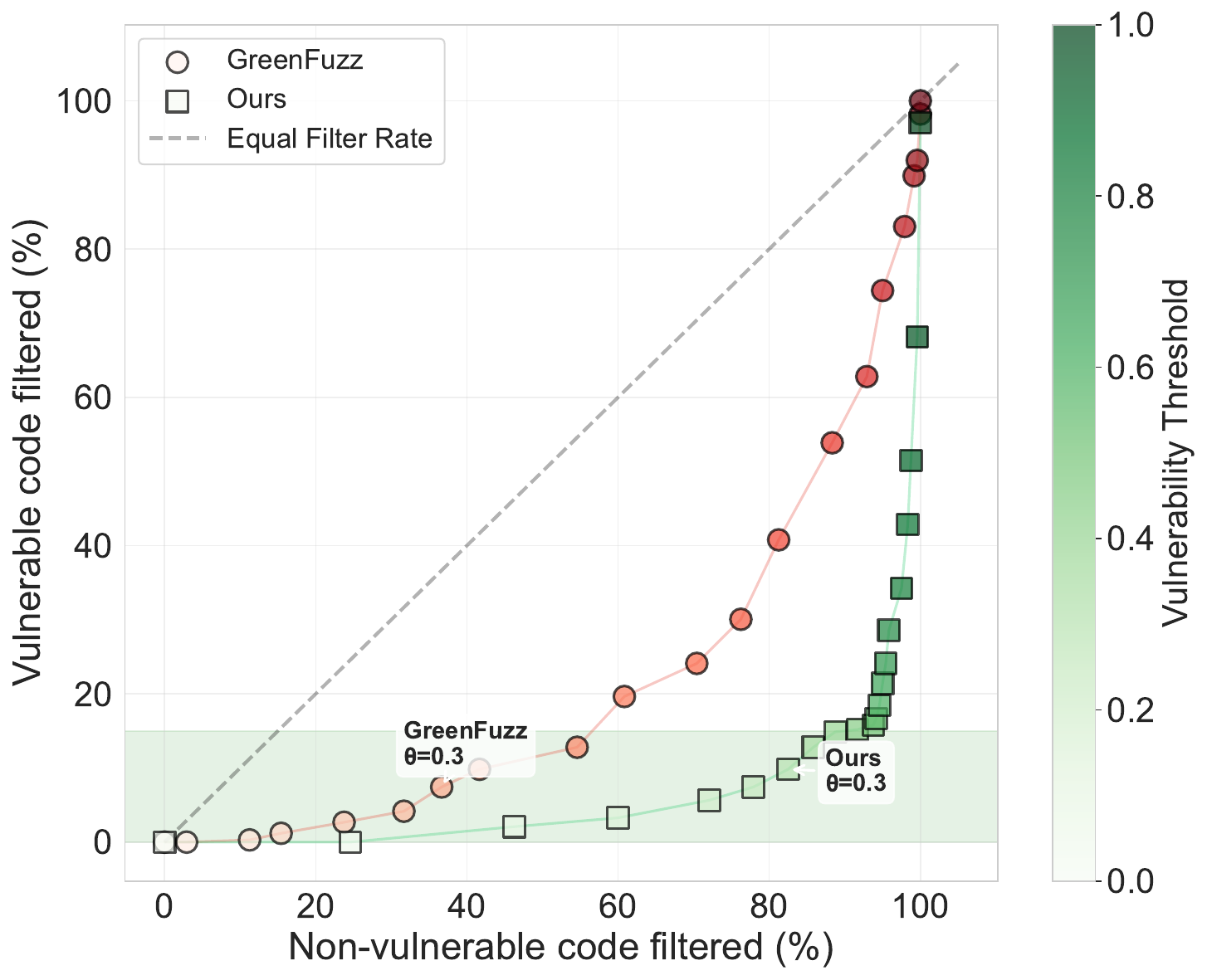}
    \caption{Comparison of discrimination capability across thresholds. \ achieves a superior trade-off by filtering significantly more non-vulnerable code (x-axis) while minimizing evicting vulnerable targets(y-axis) compared to GreenFuzz.}
    \label{fig:discrimination}
\end{figure}

The improvement can be attributed to the fact that LLM semantic features capture algorithmic complexity mismatches that static metrics miss. For example, static analysis cannot detect O($n^2$) solutions to O($n$) constrained problems, but our timeout risk feature identifies such mismatches by comparing implementation complexity against problem requirements. Similarly, our overflow risk feature can identify when a program performs long summations, and a \texttt{long} data type should have been used instead of \texttt{int}. Static analysis cannot predict whether an integer overflow will occur. However, LLMs can parse the code like a human, understand the constraints of the problem, and infer whether the input values can cause an overflow. The hybrid model correctly filters out programs that appear complex by static metrics but are algorithmically correct.

Further, to assess the generalizability of our model beyond CSES, we evaluated it on a sample of problems from the LeetCode dataset. Without any additional training, our model achieved 0.897 ROC-AUC, indicating strong transfer learning capability. However, the model's prediction ability is currently limited to algorithmic problems. Extending to general codebases or predicting common CVEs would require retraining with domain-specific vulnerability datasets.

\subsection{End-to-End Bug Detection Performance (RQ2)}

\autoref{tab:results1} shows end-to-end fuzzing performance in all approaches. 
Fixed fuzzing achieves the highest recall (81.9\%) but requires a fuzzing campaign of 336.5 minutes, which is very large considering the algorithmic problems and simply does not scale in large codebases. GreenFuzz improves over fixed fuzzing in terms of time for the fuzzing campaign by filtering out potentially non-vulnerable code. But it allocates equal resources to the chosen problems irrespective of their vulnerability score. For example, the problems \texttt{elevator\_rides\_buggy\_timeout} (buggy) and \texttt{elevator\_rides\_original} (clean) get an equal amount of fuzzing time (60s). Our hybrid approach detects 226 bugs in 141.3 minutes, achieving 1.7× speedup over GreenFuzz with comparable effectiveness. For the same set of problems, \NAME allocates 71s to \texttt{elevator\_rides\_buggy\_timeout} (buggy) and 27s to \texttt{elevator\_rides\_original} (clean) for fuzzing. The speedup gain can be justified by the usage of an improved discrimination model and efficient resource allocation.

The results demonstrate that semantic features enable intelligent prioritization, focusing fuzzing resources on high-risk targets while safely filtering likely-clean code. Overall, \NAME achieves the best time-recall tradeoff. 

\begin{table}[t]
  \centering
  \small
  \setlength{\tabcolsep}{3.5pt}
  \caption{End-to-end bug detection performance. \NAME achieves strong recall with significant time savings.}
  \label{tab:results1}
  \resizebox{\linewidth}{!}{
  \begin{tabular}{lccccc}
    \toprule
    \textbf{Approach} & \textbf{Detected} & \textbf{Filtered} & \textbf{Fuzz Time} & \textbf{Recall} & \textbf{Acc.} \\
     & \textbf{Bugs} & \textbf{Clean} & \textbf{(min)} & \textbf{(\%)} & \textbf{(\%)}  \\
    \midrule
    Fixed Fuzz & \textbf{245} & 0 & 336.5 & \textbf{72.9} & 76.7  \\
    GreenFuzz & 232 & 88 & 242.9 & 69.1 & 76.9  \\
    \NAME & 226 & \textbf{198} & \textbf{141.3} & 67.3 & \textbf{78.9}  \\
    \bottomrule
  \end{tabular}
  }
\end{table}

\subsection{Time Budget-Recall Tradeoff (RQ3)}
\begin{figure}[ht]
    \centering
    \includegraphics[width=0.8\columnwidth]{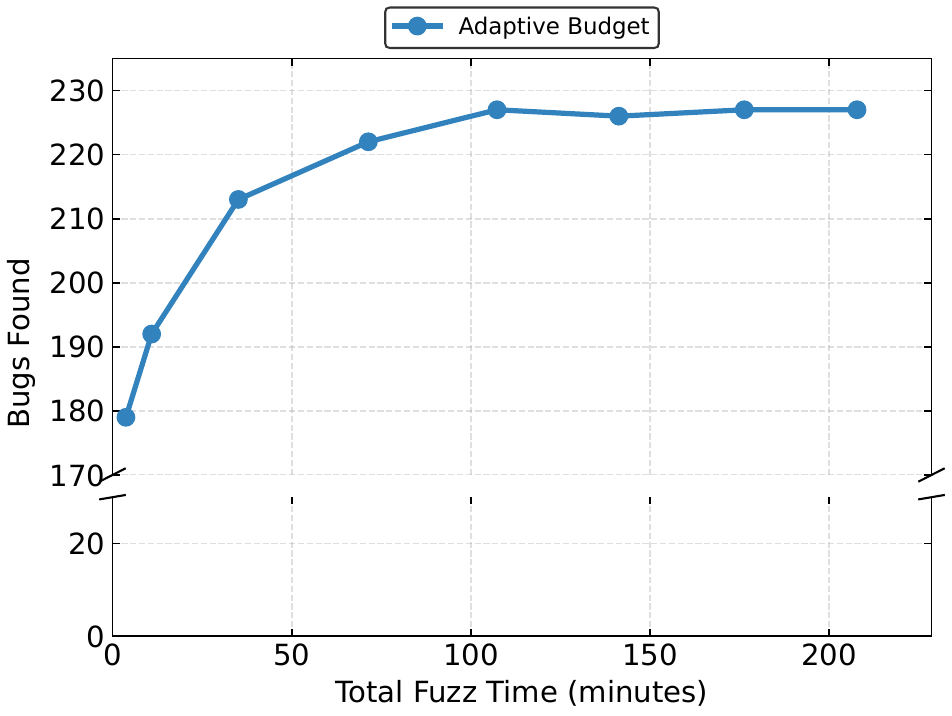}
    \caption{Tradeoff between time budget and bugs caught}
    \label{fig:time_vs_bugs}
\end{figure}

A general observation in fuzzing is that we tend to find more bugs when fuzzing for a longer time on a given harness. The question we want to answer is: how much recall do we sacrifice when choosing a specific time budget? We experimented with a range of time budgets: 2s, 5s, 15s, 30s, 45s, 60s, 75s, and 90s (time per problem), collected all true positive bugs, and showed the relationship in \autoref{fig:time_vs_bugs}. The overall trend is increasing, and the rate of increase is larger with a smaller total fuzz time, while plateauing after we go beyond 100 minutes. This experiment demonstrates that after investing a certain amount of fuzz time, the marginal gain in recall becomes minimal, providing us with valuable empirical guidelines for setting the fuzz time budget.

\subsection{Combining Fuzzing with Unit Testing (RQ4)}
As \autoref{fig:time_vs_bugs} shows, the number of detected bugs plateaus after a certain point as fuzzing time increases. A plausible explanation is that fuzzing predominantly exposes runtime failures such as crashes, timeouts, and memory-safety violations, while largely ignoring whether the program is functionally correct with respect to the problem specification. Motivated by this observation, we additionally evaluate a simple, non-integrated composition of \NAME with ChatUniTest, which focuses on functional correctness and may complement the failure-oriented nature of fuzzing. Specifically, we first run ChatUniTest on the problem set, remove programs identified as buggy, and then apply \NAME to the remaining programs.

As shown in \autoref{tab:results2}, combining the two approaches achieves an 18.1\% increase in bug detection recall compared to \NAME and a 63.8\% improvement over ChatUniTest. Moreover, since ChatUniTest filters out a substantial portion of programs upfront, the subsequent fuzzing stage requires less time, resulting in only a 4.8\% increase in runtime compared to \NAME. These results suggest that fuzzing and unit test generation tend to identify different classes of vulnerabilities, and that their combination can substantially improve vulnerability coverage.

\begin{table}[t]
  \centering
  \small
  \setlength{\tabcolsep}{3pt}
  \caption{Performance comparison among ChatUniTest, \NAME, and the combined system.}
  \label{tab:results2}
  \resizebox{\linewidth}{!}{
  \begin{tabular}{lccccc}
    \toprule
    \textbf{Approach} & \textbf{Detected} & \textbf{Filtered} & \textbf{Runtime} & \textbf{Recall} & \textbf{Acc.} \\
     & \textbf{Bugs} & \textbf{Clean} & \textbf{(min)} & \textbf{(\%)} & \textbf{(\%)}  \\
    \midrule
    ChatUniTest & 163 & 0 & \textbf{68.6} & 48.6 & 63.0  \\
    \NAME & 226 & 198 & 141.3 & 67.3 & 78.9  \\
    \NAME + ChatUniTest  & \textbf{267} & 198 & 148.1 & \textbf{79.5} & \textbf{86.1}  \\
    \bottomrule
  \end{tabular}
  }
\end{table}

\subsection{Scalability of Harness Generation Agent (RQ5)}

\begin{figure}[t]
    \centering
    \includegraphics[width=0.8\columnwidth]{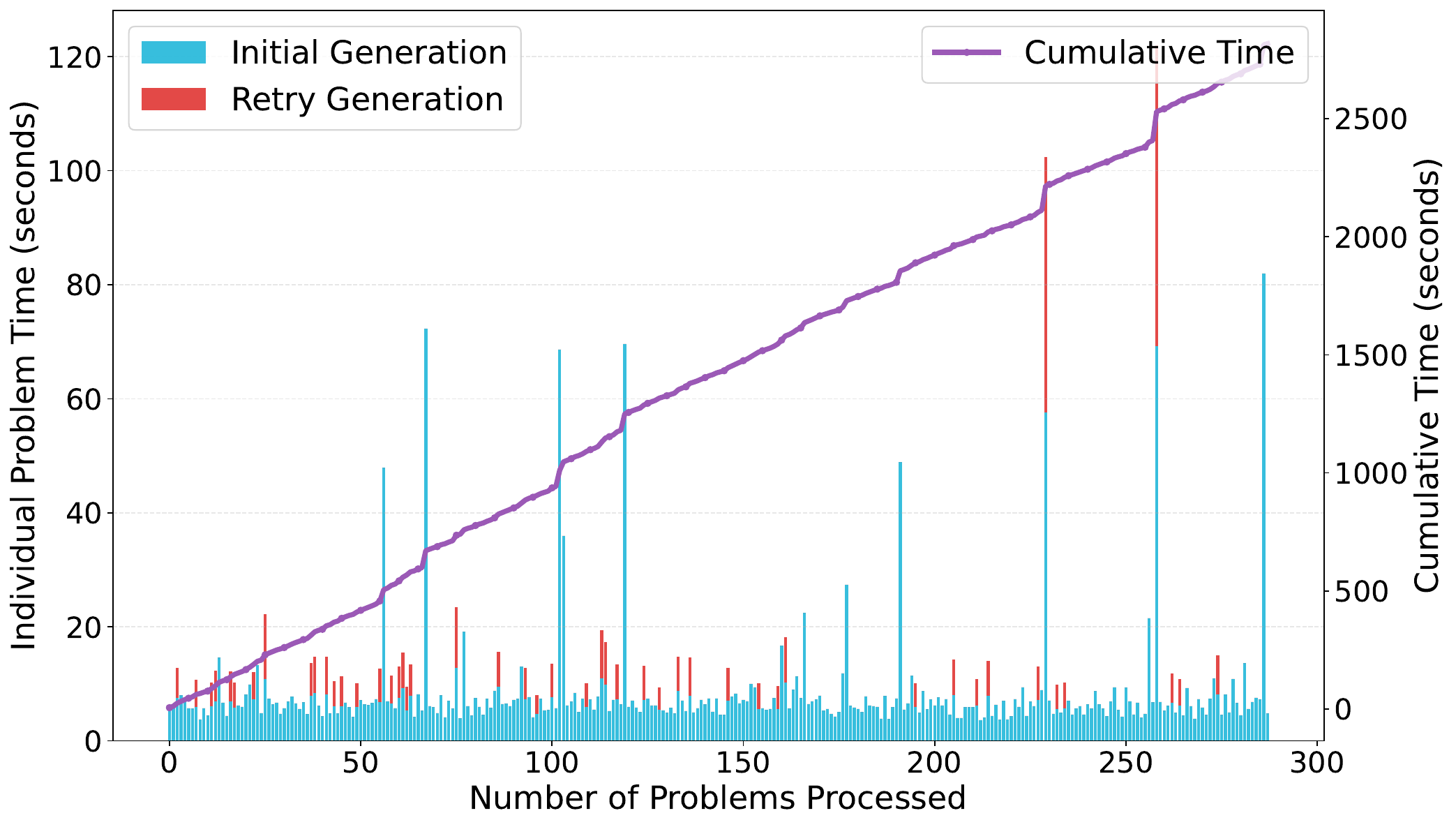}
    \caption{Assessing scalability of harness generation. The system maintains linear cost profile (purple) even when complex problems trigger retries (red spikes) ensuring the generation time grows predictably with workload size.}
    \label{fig:cumulative_fuzz_harness_gen_time}
\end{figure}

To investigate how well our fuzz harness generation agent scales, we ran experiments on 288 randomly problems selected from our test set. As \autoref{fig:cumulative_fuzz_harness_gen_time} shows, the cumulative time curve (purple) increases at a constant slope, showing that the agent scales linearly as the workload increases. We also show a breakdown of time taken per problem, where most problems spend less than 20s and either succeed at initial generation or go through the retry and fix loop. We see some time spikes corresponding to problems with higher complexity. Overall, the curve shows a good linear relation, with a few outliers but no cascading slowdowns.

\section{Related Works}
\paraf{LLM-Guided Fuzzing}
Recent work has explored using LLMs to improve fuzzing efficiency and coverage. Fuzz4All~\cite{fuzz4all2024} introduced the first universal fuzzer that uses LLMs as input generators. They employ autoprompting to encode problem descriptions into effective prompts and improve the test coverage. Although powerful, Fuzz4All is limited to testing compilers and interpreters. OSS-Fuzz-Gen~\cite{ossfuzzgen2024} automates fuzz harness generation using multi-agent LLM systems with feedback loops. However, the generated harnesses are generic and often produce false positive crashes due to incorrect input constraints. FalseCrashReducer~\cite{falsecrashreducer2025} addresses this with constraint-based drivers that reduce the number of spurious crashes. Our work differs by generating problem-specific oracles that embed semantic constraints directly from problem descriptions. As a result, we reduce false positives and catch vulnerabilities that generic harnesses miss. GreenFuzz~\cite{greenfuzz2024} proposes ML-based stopping criteria using prediction on static features, terminating fuzzing campaigns 6-12 hours earlier. 
We extend this approach with LLM semantic features that capture algorithmic properties, improving the filtering precision and better allocation of fuzzing resources.

\paraf{Agentic Code Auditing}
Recent research in AI safety has triggered a shift from static benchmarks to dynamic agent-based auditing ~\cite{redcodeagent2025, petri2025, repoaudit2025}. RepoAudit~\cite{repoaudit2025} employs LLM agents for static analysis in repository-level code auditing to detect bugs such as null pointer dereferences and memory leaks. Tools like Anthropic’s Petri~\cite{petri2025} and Microsoft’s RedCodeAgent~\cite{redcodeagent2025} utilize adversarial agent interactions to test models for unintended behaviors and security jailbreaks. They focus on invoking harmful outputs or unsafe code execution through multi-turn conversations. Our framework complements these efforts by specifically targeting algorithmic vulnerabilities through semantic-aware fuzzing and stress-testing. It fills the gap where traditional static auditing and safety red-teaming fail to detect timeout, overflow, and complexity-violation bugs.

\section{Conclusion}
In this work, we introduce \NAME, a hybrid LLM-guided fuzzing framework that detects algorithmic vulnerabilities in AI-generated code at scale. \NAME uses prompt variations to mimic user interactions, an LLM-guided fuzz harness generator that captures problem-specific constraints, generates effective oracles and stress-test inputs, and a vulnerability predictor to enable intelligent, adaptive time allocation. \NAME improves vulnerability discrimination by filtering 2$\times$ more non-vulnerable targets, achieves a 1.7$\times$ speedup compared to baselines, and enhances the bug detection recall from 67.3\% to 79.5\% when combined with LLM-based unit testing.

\balance
\bibliography{SAFuzz-Arxiv}

@misc{anthropic2024claude,
  title = {{Claude Code}: {AI}-Powered Coding Assistant},
  author = {{Anthropic}},
  year = {2025},
  howpublished = {\url{https://www.claude.com/product/claude-code}},
}

@misc{github2025copilot,
  title = {{GitHub Copilot}: Your {AI} Pair Programmer},
  author = {{GitHub}},
  year = {2025},
  howpublished = {\url{https://github.com/features/copilot}},
}

@inproceedings{chen2024chatunitest,
  title={{ChatUniTest}: A Framework for {LLM}-Based Test Generation},
  author={Chen, Yinghao and Hu, Zehao and Zhi, Chen and Han, Junxiao and Deng, Shuiguang and Yin, Jianwei},
  booktitle={Companion Proceedings of the 32nd ACM International Conference on the Foundations of Software Engineering},
  pages={572--576},
  year={2024}
}

@misc{ossfuzzgen2024,
  title = {{OSS-Fuzz-Gen}: Automated Fuzz Target Generation},
  author = {Liu, Dongge and Chang, Oliver and Metzman, Jonathan and Sablotny, Martin and Maruseac, Mihai},
  year = {2024},
  howpublished = {\url{https://github.com/google/oss-fuzz-gen}},
}

@article{manes2019art,
title = "The Art, Science, and Engineering of Fuzzing: A Survey",
author = "Manes, \{Valentin J.M.\} and Hyungseok Han and Choongwoo Han and Cha, \{Sang Kil\} and Manuel Egele and Schwartz, \{Edward J.\} and Maverick Woo",
note = "Publisher Copyright: {\textcopyright} 1976-2012 IEEE.",
year = "2021",
month = nov,
day = "1",
doi = "10.1109/TSE.2019.2946563",
language = "English",
volume = "47",
pages = "2312--2331",
journal = "IEEE Transactions on Software Engineering",
issn = "0098-5589",
number = "11",
}

@inproceedings{bohme2017coverage,
  title = {Coverage-Based Greybox Fuzzing as {Markov Chain}},
  author = {B{\"o}hme, Marcel and Pham, Van-Thuan and Roychoudhury, Abhik},
  booktitle = {Proceedings of the 2017 ACM SIGSAC Conference on Computer and Communications Security (CCS)},
  pages = {1032--1043},
  year = {2017}
}

@misc{jazzer2021,
  title = {{Jazzer}: Coverage-Guided Fuzzing for the {JVM}},
  author = {{Code Intelligence}},
  year = {2021},
  howpublished = {\url{https://github.com/CodeIntelligenceTesting/jazzer}},
}

@inproceedings{libfuzzer2016,
  title = {{libFuzzer} -- A Library for Coverage-Guided Fuzz Testing},
  author = {Serebryany, Kostya},
  booktitle = {LLVM Developers' Meeting},
  year = {2016},
  howpublished = {\url{https://llvm.org/docs/LibFuzzer.html}}
}

@inproceedings{greenfuzz2024,
author = {Lipp, Stephan and Elsner, Daniel and Kacianka, Severin and Pretschner, Alexander and B\"{o}hme, Marcel and Banescu, Sebastian},
title = {{Green Fuzzing}: A Saturation-Based Stopping Criterion using Vulnerability Prediction},
year = {2023},
isbn = {9798400702211},
publisher = {Association for Computing Machinery},
address = {New York, NY, USA},
url = {https://doi.org/10.1145/3597926.3598043},
doi = {10.1145/3597926.3598043},
booktitle = {Proceedings of the 32nd ACM SIGSOFT International Symposium on Software Testing and Analysis},
pages = {127–139},
numpages = {13},
location = {Seattle, WA, USA},
series = {ISSTA 2023}
}

@InProceedings{leino2010dafny,
author="Leino, K. Rustan M.",
editor="Clarke, Edmund M.
and Voronkov, Andrei",
title="{Dafny}: An Automatic Program Verifier for Functional Correctness",
booktitle="Logic for Programming, Artificial Intelligence, and Reasoning",
year="2010",
publisher="Springer Berlin Heidelberg",
address="Berlin, Heidelberg",
pages="348--370",
isbn="978-3-642-17511-4"
}

@misc{akhond2025llmloopinvariant,
      title={{LLM} For Loop Invariant Generation and Fixing: How Far Are We?}, 
      author={Mostafijur Rahman Akhond and Saikat Chakraborty and Gias Uddin},
      year={2025},
      eprint={2511.06552},
      archivePrefix={arXiv},
      primaryClass={cs.SE},
      url={https://arxiv.org/abs/2511.06552}
}

@misc{wei2025invbench,
      title={{InvBench}: Can {LLMs} Accelerate Program Verification with Invariant Synthesis?}, 
      author={Anjiang Wei and Zhiyuan Yan and Hanjun Dai and Yuxiang Wei and Jingxuan He and Maolin Wei and Xujie Si},
      year={2025},
      eprint={2509.21629},
      archivePrefix={arXiv},
      primaryClass={cs.SE},
      url={https://arxiv.org/abs/2509.21629}
}

@misc{understand2025,
  title = {{SciTools Understand}: Static Code Analysis Tool},
  author = {{SciTools}},
  year = {2025},
  howpublished = {\url{https://scitools.com/}},
}

@misc{sarker2024syntacticrobustnessllmbasedcode,
      title={Syntactic Robustness for {LLM}-based Code Generation}, 
      author={Laboni Sarker and Mara Downing and Achintya Desai and Tevfik Bultan},
      year={2024},
      eprint={2404.01535},
      archivePrefix={arXiv},
      primaryClass={cs.SE},
      url={https://arxiv.org/abs/2404.01535}, 
}

@misc{paleyes2025promptvariabilityeffectsllm,
      title={Prompt Variability Effects On {LLM} Code Generation}, 
      author={Andrei Paleyes and Radzim Sendyka and Diana Robinson and Christian Cabrera and Neil D. Lawrence},
      year={2025},
      eprint={2506.10204},
      archivePrefix={arXiv},
      primaryClass={cs.SE},
      url={https://arxiv.org/abs/2506.10204}, 
}

@article{chen2025nlperturbator,
author = {Chen, Junkai and Zhenhao, Li and Xing, Hu and Xin, Xia},
title = {{NLPerturbator}: Studying the Robustness of Code {LLMs} to Natural Language Variations},
year = {2025},
publisher = {Association for Computing Machinery},
address = {New York, NY, USA},
issn = {1049-331X},
url = {https://doi.org/10.1145/3745764},
doi = {10.1145/3745764},
note = {Just Accepted},
journal = {ACM Trans. Softw. Eng. Methodol.},
month = jul
}

@misc{cses2025,
  title = {{CSES} Problem Set},
  author = {Laaksonen, Antti},
  year = {2025},
  howpublished = {\url{https://cses.fi/problemset/}},
}

@misc{deepinfra2025,
  title = {{DeepInfra}: {LLM} Models and Infrastructure},
  author = {{DeepInfra}},
  year = {2025},
  howpublished = {\url{https://deepinfra.com/}},
}

@misc{qwen3technicalreport,
  title = {{Qwen3} Technical Report},
  author = {{Qwen Team}},
  year = {2025},
  eprint = {2505.09388},
  archivePrefix = {arXiv},
  primaryClass = {cs.CL},
  url = {https://arxiv.org/abs/2505.09388}
}

@article{breiman2001random,
author = {Breiman, Leo},
title = {{Random Forests}},
year = {2001},
issue_date = {October 1 2001},
publisher = {Kluwer Academic Publishers},
address = {USA},
volume = {45},
number = {1},
issn = {0885-6125},
url = {https://doi.org/10.1023/A:1010933404324},
doi = {10.1023/A:1010933404324},
journal = {Mach. Learn.},
month = oct,
pages = {5–32},
numpages = {28}
}

@article{pedregosa2011scikit,
  title={{Scikit-learn}: Machine Learning in {Python}},
  author={Pedregosa, F. and Varoquaux, G. and Gramfort, A. and Michel, V.
          and Thirion, B. and Grisel, O. and Blondel, M. and Prettenhofer, P.
          and Weiss, R. and Dubourg, V. and Vanderplas, J. and Passos, A. and
          Cournapeau, D. and Brucher, M. and Perrot, M. and Duchesnay, E.},
  journal={Journal of Machine Learning Research},
  volume={12},
  pages={2825--2830},
  year={2011}
}

@inproceedings{repoaudit2025,
  title={{RepoAudit}: An Autonomous {LLM}-Agent for Repository-Level Code Auditing},
  author={Guo, Jinyao* and Wang, Chengpeng* and Xu, Xiangzhe and Su, Zian and Zhang, Xiangyu},
  booktitle={Proceedings of the 42nd International Conference on Machine Learning},
  year={2025},
  note={*Equal contribution}
}

@inproceedings{fuzz4all2024,
  author = {Xia, Chunqiu Steven and Paltenghi, Matteo and Tian, Jia Le and Pradel, Michael and Zhang, Lingming},
  title = {{Fuzz4All}: Universal Fuzzing with Large Language Models},
  booktitle = {Proceedings of the IEEE/ACM International Conference on Software Engineering (ICSE)},
  year = {2024}
}

@misc{falsecrashreducer2025,
  title={{FalseCrashReducer}: Mitigating False Positive Crashes in {OSS-Fuzz-Gen} Using Agentic {AI}},
  author={Amusuo, Paschal C. and Liu, Dongge and Calvo Méndez, Ricardo Andrés and Metzman, Jonathan and Chang, Oliver and Davis, James C.},
  year={2025},
  eprint={2510.02185},
  archivePrefix={arXiv},
  primaryClass={cs.SE},
  url={https://arxiv.org/abs/2510.02185}
}

@misc{redcodeagent2025,
  title={{RedCodeAgent}: Automatic Red-teaming Agent against Diverse Code Agents},
  author={Guo, Chengquan and Xie, Chulin and Yang, Yu and Chen, Zhaorun and Lin, Zinan and Davies, Xander and Gal, Yarin and Song, Dawn and Li, Bo},
  year={2025},
  eprint={2510.02609},
  archivePrefix={arXiv},
  primaryClass={cs.SE},
  url={https://arxiv.org/abs/2510.02609}
}

@misc{petri2025,
  title={{Petri}: Parallel Exploration of Risky Interactions},
  author={Fronsdal, Kai and Gupta, Isha and Sheshadri, Abhay and Michala, Jonathan and McAleer, Stephen and Wang, Rowan and Price, Sara and Bowman, Sam},
  year={2025},
  url={https://github.com/safety-research/petri},
}

@inproceedings{bohme2020fuzzing,
  title = {Fuzzing: On the Exponential Cost of Vulnerability Discovery},
  author = {B{\"o}hme, Marcel and Falk, Brandon},
  booktitle = {Proceedings of the 28th ACM Joint Meeting on European Software Engineering Conference and Symposium on the Foundations of Software Engineering (ESEC/FSE)},
  pages = {713--724},
  year = {2020},
  doi = {10.1145/3368089.3409729}
}

@inproceedings{afl2020,
author = {Fioraldi, Andrea and Maier, Dominik and Ei\ss{}feldt, Heiko and Heuse, Marc},
title = {{AFL++}: combining incremental steps of fuzzing research},
year = {2020},
publisher = {USENIX Association},
address = {USA},
booktitle = {Proceedings of the 14th USENIX Conference on Offensive Technologies},
articleno = {10},
numpages = {1},
series = {WOOT'20}
}

@misc{yue2025surveylargelanguagemodel,
      title={A Survey of Large Language Model Agents for Question Answering}, 
      author={Murong Yue},
      year={2025},
      eprint={2503.19213},
      archivePrefix={arXiv},
      primaryClass={cs.CL},
      url={https://arxiv.org/abs/2503.19213}, 
}

@article{Guo_2025,
   title={Deep{S}eek-{R1} incentivizes reasoning in {LLMs} through reinforcement learning},
   volume={645},
   ISSN={1476-4687},
   url={http://dx.doi.org/10.1038/s41586-025-09422-z},
   DOI={10.1038/s41586-025-09422-z},
   number={8081},
   journal={Nature},
   publisher={Springer Science and Business Media LLC},
   author={Guo, Daya and Yang, Dejian and Zhang, Haowei and Song, Junxiao and Wang, Peiyi and Zhu, Qihao and Xu, Runxin and Zhang, Ruoyu and Ma, Shirong and Bi, Xiao and Zhang, Xiaokang and Yu, Xingkai and Wu, Yu and Wu, Z. F. and Gou, Zhibin and Shao, Zhihong and Li, Zhuoshu and Gao, Ziyi and Liu, Aixin and Xue, Bing and Wang, Bingxuan and Wu, Bochao and Feng, Bei and Lu, Chengda and Zhao, Chenggang and Deng, Chengqi and Ruan, Chong and Dai, Damai and Chen, Deli and Ji, Dongjie and Li, Erhang and Lin, Fangyun and Dai, Fucong and Luo, Fuli and Hao, Guangbo and Chen, Guanting and Li, Guowei and Zhang, H. and Xu, Hanwei and Ding, Honghui and Gao, Huazuo and Qu, Hui and Li, Hui and Guo, Jianzhong and Li, Jiashi and Chen, Jingchang and Yuan, Jingyang and Tu, Jinhao and Qiu, Junjie and Li, Junlong and Cai, J. L. and Ni, Jiaqi and Liang, Jian and Chen, Jin and Dong, Kai and Hu, Kai and You, Kaichao and Gao, Kaige and Guan, Kang and Huang, Kexin and Yu, Kuai and Wang, Lean and Zhang, Lecong and Zhao, Liang and Wang, Litong and Zhang, Liyue and Xu, Lei and Xia, Leyi and Zhang, Mingchuan and Zhang, Minghua and Tang, Minghui and Zhou, Mingxu and Li, Meng and Wang, Miaojun and Li, Mingming and Tian, Ning and Huang, Panpan and Zhang, Peng and Wang, Qiancheng and Chen, Qinyu and Du, Qiushi and Ge, Ruiqi and Zhang, Ruisong and Pan, Ruizhe and Wang, Runji and Chen, R. J. and Jin, R. L. and Chen, Ruyi and Lu, Shanghao and Zhou, Shangyan and Chen, Shanhuang and Ye, Shengfeng and Wang, Shiyu and Yu, Shuiping and Zhou, Shunfeng and Pan, Shuting and Li, S. S. and Zhou, Shuang and Wu, Shaoqing and Yun, Tao and Pei, Tian and Sun, Tianyu and Wang, T. and Zeng, Wangding and Liu, Wen and Liang, Wenfeng and Gao, Wenjun and Yu, Wenqin and Zhang, Wentao and Xiao, W. L. and An, Wei and Liu, Xiaodong and Wang, Xiaohan and Chen, Xiaokang and Nie, Xiaotao and Cheng, Xin and Liu, Xin and Xie, Xin and Liu, Xingchao and Yang, Xinyu and Li, Xinyuan and Su, Xuecheng and Lin, Xuheng and Li, X. Q. and Jin, Xiangyue and Shen, Xiaojin and Chen, Xiaosha and Sun, Xiaowen and Wang, Xiaoxiang and Song, Xinnan and Zhou, Xinyi and Wang, Xianzu and Shan, Xinxia and Li, Y. K. and Wang, Y. Q. and Wei, Y. X. and Zhang, Yang and Xu, Yanhong and Li, Yao and Zhao, Yao and Sun, Yaofeng and Wang, Yaohui and Yu, Yi and Zhang, Yichao and Shi, Yifan and Xiong, Yiliang and He, Ying and Piao, Yishi and Wang, Yisong and Tan, Yixuan and Ma, Yiyang and Liu, Yiyuan and Guo, Yongqiang and Ou, Yuan and Wang, Yuduan and Gong, Yue and Zou, Yuheng and He, Yujia and Xiong, Yunfan and Luo, Yuxiang and You, Yuxiang and Liu, Yuxuan and Zhou, Yuyang and Zhu, Y. X. and Huang, Yanping and Li, Yaohui and Zheng, Yi and Zhu, Yuchen and Ma, Yunxian and Tang, Ying and Zha, Yukun and Yan, Yuting and Ren, Z. Z. and Ren, Zehui and Sha, Zhangli and Fu, Zhe and Xu, Zhean and Xie, Zhenda and Zhang, Zhengyan and Hao, Zhewen and Ma, Zhicheng and Yan, Zhigang and Wu, Zhiyu and Gu, Zihui and Zhu, Zijia and Liu, Zijun and Li, Zilin and Xie, Ziwei and Song, Ziyang and Pan, Zizheng and Huang, Zhen and Xu, Zhipeng and Zhang, Zhongyu and Zhang, Zhen},
   year={2025},
   month=sep, pages={633–638} }

@misc{jiang2025ragosystematicperformanceoptimization,
      title={{RAGO}: Systematic Performance Optimization for Retrieval-Augmented Generation Serving}, 
      author={Wenqi Jiang and Suvinay Subramanian and Cat Graves and Gustavo Alonso and Amir Yazdanbakhsh and Vidushi Dadu},
      year={2025},
      eprint={2503.14649},
      archivePrefix={arXiv},
      primaryClass={cs.IR},
      url={https://arxiv.org/abs/2503.14649}, 
}

@misc{zheng2025automationautonomysurveylarge,
      title={From Automation to Autonomy: A Survey on Large Language Models in Scientific Discovery}, 
      author={Tianshi Zheng and Zheye Deng and Hong Ting Tsang and Weiqi Wang and Jiaxin Bai and Zihao Wang and Yangqiu Song},
      year={2025},
      eprint={2505.13259},
      archivePrefix={arXiv},
      primaryClass={cs.CL},
      url={https://arxiv.org/abs/2505.13259}, 
}

@article{Abramson2024,
  author  = {Abramson, Josh and Adler, Jonas and Dunger, Jack and Evans, Richard and Green, Tim and Pritzel, Alexander and Ronneberger, Olaf and Willmore, Lindsay and Ballard, Andrew J. and Bambrick, Joshua and Bodenstein, Sebastian W. and Evans, David A. and Hung, Chia-Chun and O’Neill, Michael and Reiman, David and Tunyasuvunakool, Kathryn and Wu, Zachary and Žemgulytė, Akvilė and Arvaniti, Eirini and Beattie, Charles and Bertolli, Ottavia and Bridgland, Alex and Cherepanov, Alexey and Congreve, Miles and Cowen-Rivers, Alexander I. and Cowie, Andrew and Figurnov, Michael and Fuchs, Fabian B. and Gladman, Hannah and Jain, Rishub and Khan, Yousuf A. and Low, Caroline M. R. and Perlin, Kuba and Potapenko, Anna and Savy, Pascal and Singh, Sukhdeep and Stecula, Adrian and Thillaisundaram, Ashok and Tong, Catherine and Yakneen, Sergei and Zhong, Ellen D. and Zielinski, Michal and Žídek, Augustin and Bapst, Victor and Kohli, Pushmeet and Jaderberg, Max and Hassabis, Demis and Jumper, John M.},
  journal = {Nature},
  title   = {Accurate structure prediction of biomolecular interactions with AlphaFold 3},
  year    = {2024},
  volume  = {630},
  number  = {8016},
  pages   = {493–-500},
  doi     = {10.1038/s41586-024-07487-w}
}

@software{cursor_editor,
  author = {{Anysphere}},
  title = {Cursor: The {AI}-First Code Editor},
  url = {https://www.cursor.com},
  year = {2023},
  note = {Accessed: 2026-01-28}
}

@article{productivityllm,
author = {Weber, Thomas and Brandmaier, Maximilian and Schmidt, Albrecht and Mayer, Sven},
title = {Significant Productivity Gains through Programming with Large Language Models},
year = {2024},
issue_date = {June 2024},
publisher = {Association for Computing Machinery},
address = {New York, NY, USA},
volume = {8},
number = {EICS},
url = {https://doi.org/10.1145/3661145},
doi = {10.1145/3661145},
journal = {Proc. ACM Hum.-Comput. Interact.},
month = jun,
articleno = {256},
numpages = {29}
}

@inproceedings{jimenez2024swebench,
  title     = {SWE-bench: Can Language Models Resolve Real-World {G}it{H}ub Issues?},
  author    = {Jimenez, Carlos E. and Yang, John and Wettig, Alexander and Yao, Shunyu and Pei, Kexin and Press, Ofir and Narasimhan, Karthik},
  booktitle = {International Conference on Learning Representations (ICLR)},
  year      = {2024}
}

@inproceedings{pearce2022copilot,
  title     = {Asleep at the Keyboard? Assessing the Security of {G}it{H}ub {C}opilot's Code Contributions},
  author    = {Pearce, Hammond and Ahmad, Baleegh and Tan, Benjamin and Dolan-Gavitt, Brendan and Karri, Ramesh},
  booktitle = {2022 {IEEE} Symposium on Security and Privacy ({SP})},
  year      = {2022},
  pages     = {754--768}
}

@article{manes2021fuzzing,
  title   = {The Art, Science, and Engineering of Fuzzing: A Survey},
  author  = {Man{\`e}s, Valentin J. M. and Han, HyungSeok and Han, Choongwoo and Cha, Sang Kil and Egele, Manuel and Schwartz, Edward J. and Woo, Maverick},
  journal = {{IEEE} Transactions on Software Engineering},
  year    = {2021},
  volume  = {47},
  number  = {11},
  pages   = {2312--2331},
  doi     = {10.1109/TSE.2019.2946563}
}

@inproceedings{liu2024promptinj,
  title     = {Formalizing and Benchmarking Prompt Injection Attacks and Defenses},
  author    = {Liu, Yupei and Jia, Yuqi and Geng, Runpeng and Jia, Jinyuan and Gong, Neil Zhenqiang},
  booktitle = {33rd {USENIX} Security Symposium ({USENIX} Security)},
  year      = {2024},
  pages     = {1831--1847}
}

@misc{cotroneo2025humanai,
  title={Human-Written vs. AI-Generated Code: A Large-Scale Study of Defects, Vulnerabilities, and Complexity},
  author={Cotroneo, Domenico and Improta, Cristina and Liguori, Pietro},
  year={2025},
  eprint={2508.21634},
  archivePrefix={arXiv},
  primaryClass={cs.SE},
  url={https://arxiv.org/abs/2508.21634}, 
}

@article{quantumrun2026copilot,
  title={GitHub Copilot Statistics 2026: User Growth and Adoption Trends},
  author={{Quantumrun Foresight}},
  journal={Quantumrun Consulting Reports},
  year={2026},
  url={https://www.quantumrun.com/consulting/github-copilot-statistics/}
}

@misc{burke2025robotbuildsrobotsbrain,
      title={Robot builds a robot's brain: AI generated drone command and control station hosted in the sky}, 
      author={Peter Burke},
      year={2025},
      eprint={2508.02962},
      archivePrefix={arXiv},
      primaryClass={cs.RO},
      url={https://arxiv.org/abs/2508.02962}, 
}

@inproceedings{polo,
author = {Bai, Jiameng and Xu, Ruoyi and Wu, Sai and Yang, Dingyu and Zhao, Junbo and Chen, Gang},
title = {POLO: an LLM-powered project-level code performance optimization framework},
year = {2025},
isbn = {978-1-956792-06-5},
url = {https://doi.org/10.24963/ijcai.2025/814},
doi = {10.24963/ijcai.2025/814},
booktitle = {Proceedings of the Thirty-Fourth International Joint Conference on Artificial Intelligence},
articleno = {814},
numpages = {10},
location = {Montreal, Canada},
series = {IJCAI '25}
}

@article{hlsrewriter,
author = {Xu, Kangwei and Zhang, Grace Li and Yin, Xunzhao and Zhuo, Cheng and Schlichtmann, Ulf and Li, Bing},
title = {HLSRewriter: Efficient Refactoring and Optimization of C/C++ Code with LLMs for High-Level Synthesis},
year = {2025},
publisher = {Association for Computing Machinery},
address = {New York, NY, USA},
issn = {1084-4309},
url = {https://doi.org/10.1145/3749986},
doi = {10.1145/3749986},
note = {Just Accepted},
journal = {ACM Trans. Des. Autom. Electron. Syst.},
month = jul
}
\bibliographystyle{icml2026}

\end{document}